\documentclass[10pt,conference]{IEEEtran}
\usepackage{amsmath,amssymb,amsthm,textcomp,bm, amsfonts, amscd,latexsym,wasysym,mathrsfs,bbm}
\usepackage{setspace}
\usepackage{enumerate}
\usepackage{subfigure}
\usepackage[noadjust]{cite}
\usepackage{epsfig}
\usepackage{multirow}
\usepackage{refstyle}

\usepackage{mdwmath}
\usepackage{mdwtab}

\newtheorem{thm1}{Theorem}[section]

\newtheorem{lem}{Lemma}[section]

\theoremstyle{definition}

\newtheorem{eg1}{Example}[section]

\numberwithin{equation}{section}
\numberwithin{figure}{section}
\numberwithin{table}{section}

\def\epr{\text{P}_{\text{err}}^{\text{ge}}}
\def\mcd{\mathcal{D}}

\makeatletter
\def\cdotfill{%
  \leavevmode
  \cleaders \hb@xt@ .44em{\hss$\cdot$\hss}\hfill
  \kern\z@}
\makeatother

\newcommand{\sysref}[1]{system~(\ref{#1})}
\renewcommand{\secref}[1]{Section~\ref{#1}}

\renewcommand{\figref}[1]{Fig.~\ref{#1}}
\renewcommand{\tabref}[1]{{\sc table}~\ref{#1}}

\begin{document}

\title {A Tight Estimate for Decoding Error-Probability of  LT Codes Using  Kovalenko's Rank Distribution} 

\author{
\authorblockN{Ki-Moon Lee}
\authorblockA{Dept of Info. \& Comm. Eng\\
Sungkyunkwan University \\
Suwon Kyunggi, R.O.Korea\\
Email: leekimoo@skku.edu}
\and
\authorblockN{Hayder Radha}
\authorblockA{Department of ECE\\
Michigan State University\\
E. Lansing, MI 48824, USA\\
Email: radha@egr.msu.edu}
\and
\authorblockN{Beom-Jin Kim
}
\authorblockA{Department of Mathematics \\
Yonsei University\\
Seoul, R.O.Korea\\
Email: beomjinkim@yonsei.ac.kr
}
}

\maketitle

\begin{abstract}
A new approach for estimating the Decoding Error-Probability (DEP) of LT codes with dense rows is derived by using  the  conditional Kovalenko's rank distribution.  
The  estimate by the proposed approach  is very close to the DEP approximated by Gaussian Elimination, and is significantly less complex. 
As  a key application, we utilize the estimates for obtaining optimal LT codes with dense rows, whose DEP is very close to the  Kovalenko's Full-Rank Limit within a desired error-bound.
Experimental evidences which show the viability of the estimates are also provided.
\end{abstract}
\section{Introduction and Backgrounds}

For Binary Erasure Channels (BEC), the task of a Luby Transform (LT) decoder is to recover the unique solution of  a consistent linear system
\begin{equation}
HX^T=\beta^T, \quad \beta=(\beta_1,\dots,\beta_m)\in (\mathbb{F}_2^s)^m, \label{sys:ini}
\end{equation} 
where $H$ is an $m\times n$ matrix over $\mathbb{F}_2$.  
This can be explained briefly as follows. (For detailed backgrounds, see \cite{lt,raptor,mackay2}).
In LT codes, to communicate an information symbol vector  $\alpha=(\alpha_1,\dots,\alpha_n) \in (\mathbb{F}_2^s)^n$, a sender constantly generates and transmits  a syndrome symbol  $\beta_i=H_i\alpha^T$ over BEC, where   $H_i\in \mathbb{F}_2^n$ is generated uniformly at random  on the fly by using the Robust Soliton Distribution (RSD) $\mu(x)$  \cite{lt}.   
A receiver then acquires  a set of pairs $\{ (H_{i_t},\beta_{i_t})\}_{t=1}^{m}$ and interprets it as \sysref{sys:ini}, and hence, the variable vector  $X=(x_1,\dots,x_n)\in (\mathbb{F}_2^s)^n$  represents the information symbol vector   $\alpha$. Unlike LDPC codes, the row-dimension $m$ of $H$ is a variable and the column-dimension $n$  is fixed. Thus a reception overhead defined as $\gamma=\frac{m-n}{n}$ is the key parameter for measuring error-performance of LT codes.

System~(\ref{sys:ini}) has the unique solution $X=\alpha$ iff $\text{Rank}(H)=n$, the full rank of $H$. In case of  the full-rank, $\alpha$ can be recovered  by using a Maximum-Likelihood Decoding Algorithm (MLDA) such as the ones in \cite{bnm1,oltdr}.  
These algorithms are an efficient  Gaussian Elimination (GE) that fully utilize an approximate lower triangulation of $H$,  obtainable by exploiting the diagonal extension process with various greedy algorithms in \cite{eeldpc,bnm1,usp1,oltdr}.   
  Under those GE, thus, the probability of decoding success is  the full-rank probability  $\Pr(\text{Rank}(H)=n)$.
  
  It is shown in \cite{lt} by Luby that, when $H$ is  generated by the RSD with large $n$ and $\gamma \ge \ln(n/\delta)/\sqrt{n}$, \sysref{sys:ini} can be solved for $X=\alpha$  by using the Message Passing Algorithm (MPA) in \cite{tornado1} with the minimum probability $1-\delta$, and the number of row operations to compute $X=\alpha$ by the MPA is  $O(n\ln(n/\delta))$.
For short $n$, however, a stable overhead $\gamma$ needed for successful decoding by the MPA with high probability  is not trivial.  In fact, even under GE that is much superior to the MPA in error-performance, a stable $\gamma$ to achieve the full-rank probability near one is not trivial.

Let $H$ of \sysref{sys:ini} be an $m\times n$ binary random matrix  generated by a row-degree distribution $\rho(x)=\sum \rho_d x^d$ with $m=(1+\gamma)n$.
The Decoding Error Probability (DEP) of an LT code generated by $\rho(x)$ used in this paper is the rank-deficient probability defined as
\begin{eqnarray}
\epr (1+\gamma,n,\rho)&=&1-\Pr(\text{Rank}(H)=n) \label{eq:dfr}\\
                      &=&\Pr(\text{Rank}(H)<n). \label{eq:dfr2}
\end{eqnarray}
Then with a desired error-bound $\delta\in(0,1)$,  define 
\begin{equation}
\gamma_{\min}(\delta,n,\rho) = \min_{\gamma \ge 0}\{ \gamma\: |\: \epr (1+\gamma,n,\rho)\le\delta \}, \label{eq:mo}
\end{equation}
and refer to as the Minimum Stable Overhead (MSO) of the code with $\delta$.
With  $m=(1+\gamma)n$ symbols of $\beta$ where $\gamma \ge \gamma_{\min}(\delta,n,\rho)$, thus, the recovery of $\alpha$ can be accomplished by GE decoding with probability at least $1-\delta$. 

It was shown in \cite{kfrl}  that probabilistic lower-bounds for  DEP and MSO of random binary codes exist, called Kovalenko's Full-Rank Limit and Overhead (KFRL and KFRO respectively). Specifically, KFRL is the function
\begin{equation}
K(1+\gamma,n) = 1-\prod_{i=k+1}^n\left(1-\frac{1}{2^i}\right),\quad k=\gamma n, \label{eq:kfrl0}
\end{equation}
where $K(1+\gamma,n)\le \epr(1+\gamma,n,\rho)$.  
 Hence, the DEP of LT and LDPC codes  cannot be lower than  KFRL. 
Similar to MSO, KFRO  is  the minimum  $\gamma$ defined as
\begin{equation}
\gamma_K(\delta,n)=\min_{\gamma}\{\gamma \ge 0 \: | \: K(1+\gamma,n)\le \delta  \}, \label{eq:kfro0}
\end{equation}
where $\gamma_K(\delta,n) \le \gamma_{\min}(\delta,n,\rho)$.
For successful decoding under the constraint $\epr(1+\gamma, n, \rho)\le \delta$, thus,  the minimum number of symbols of $\beta$ that a receiver should acquire is at least $(1+\gamma_K(\delta,n))n$ \cite[{\bf Theorem~2.2}]{kfrl}.
For short $n$ and small $\delta$, since the KFRO $\gamma_K(\delta,n)$ is not trivial, the MSO $\gamma_{\min}(\delta,n,\rho)$ is not trivial. 
It was also observed experimentally that $K(1+\gamma,n)\approx 2^{-\gamma n}$ as $\gamma$ increases. For small $\delta$, therefore, $\gamma_K(\delta,n) \approx \frac{k_\delta}{n}$, where $k_\delta=\min\{k  \in \mathbb{N}^+\;|\;  2^{-k} < \delta\}$.

Let $\mu(x)=\sum_{d\le d_0}\mu_dx^d$, a truncated form of RSD where $\lim_{n\to\infty}\frac{d_0}{n}=0$.
For short $n$, the DEP of codes (generated) by a truncated $\mu(x)$ alone exhibits error-floors over a large range of $\gamma$.  The error-floor region however can be lowered   to near zero dramatically by supplementing  a small fraction of dense rows to $\mu(x)$ (see \cite{isit07,allerton07,kfrl}).
The row-degree distribution $\rho(x)$ considered in this paper is thus a supplementation of  $\mu(x)$ with a fraction of rows of degree $\frac{n}{2}$ as shown below
\begin{equation}
\rho(x)=\sum_{d\le d_0} \left(\frac{\mu_d}{1+\kappa}\right) x^d + \left(\frac{\kappa}{1+\kappa}\right)x^{n/2},\quad \kappa \ge 0 . \label{eq:rho1}
\end{equation}
By rearranging rows, an $H$ by $\rho(x)$ above can be expressed as $H=\left[\begin{smallmatrix}\tilde{H} \\ \hat{H} \end{smallmatrix} \right]$, where $\tilde{H}$ is a sparse matrix generated by $\mu(x)$ and $\hat{H}$ is a dense one formed by random rows of degree $\frac{n}{2}$.
The key objective of optimizing LT codes  in this paper is to obtain a $\rho(x)$, by which, generated LT codes can achieve  the $\gamma_{\min}(\delta,n,\rho)$ near the $\gamma_K(\delta, n)$ for better error-performance, but the dense fraction $\rho_{n/2}=\frac{\kappa}{1+\kappa}$  is as small as possible for encoding and decoding efficiency. 


In the paper \cite{oltdr}, a simple way of using an Upper-Bound of DEP (UBDEP) was formulated for the fast optimization.  This approach was quite effective in that, an estimate of the UBDEP by the formulation is close to the  DEP approximated by GE  and is obtainable very rapidly (within a fraction of a second using a standard computer). 
Hence, the optimization was accomplished very rapidly as well  by checking the estimates with  various fractions for $\rho_{n/2}$.

In this paper, an exact formulation of DEP is derived by decomposing the full-rank probability in (\ref{eq:dfr})   as a sum of conditional full-rank probabilities, that are computable quite rapidly by using  by  the conditional Kovalenko's rank-distribution. 
The formulation  is similar to that of UBDEP  in  that, it uses  prior knowledges of the rank-distribution of the sparse part $\tilde{H}$.  
The Estimate of DEP (EDEP) by the formulation is however extremely close to the DEP approximated  by  GE, and also  is computable quite rapidly (again, within a fraction of a second on a standard computer).
Thus, a finer optimization of $\rho(x)$ can be accomplished very fast  by checking  EDEP's with  various fractions for $\rho_{n/2}$.

The remainder of this paper is organized as follows.
In \secref{trsd}, a simple approach for generating a truncated RSD for a supplementation $\rho(x)$ in (\ref{eq:rho1}) is presented. 
In \secref{testimate}, explicit formulations of DEP and UBDEP are derived  by analyzing the full-rank probability in (\ref{eq:dfr}) and the rank-deficient probability  in (\ref{eq:dfr2}), respectively, and are utilized for obtaining an optimal $\rho(x)$.
The KFRL in (\ref{eq:kfrl0})  is also explained by the conditional Kovalenko's rank-distribution in this section.
In \secref{simulation},  further experimental results  which show the viability of the EDEP  are presented.  
This paper is  summarized  in \secref{conclusion}.

\section{ A Truncation of an RSD \label{trsd}}

The RSD considered for the truncation  in (\ref{eq:rho1}) is the one in \cite[Ex.50.2]{mackay2}.
Let $h(d)$ denote  the expected number of rows of degree $d$ of an $H$  by $\mu(x)$.  
With $S\ge 1$, we have
\begin{equation}
 h(d)=
  \begin{cases}
    S+1, & \text{if } d=1 \\
    \frac{n}{d(d-1)} + \frac{S}{d}, &\text{otherwise} \label{h0d}
  \end{cases}.
\end{equation}
Setting  the number of rows of $H$ as
\begin{equation}
m=\sum_{d=1}^n h(d) \approx n+ S(1+\ln(n)), \label{eq:numr}
\end{equation}
then normalizing $h(d)$ by the $m$ yields 
\begin{equation}
 \mu(x)=\sum_{d=1}^n \mu_d x^d, \quad \text{where} \quad \mu_d=\frac{h(d)}{m}. \label{eq:grsd} 
\end{equation}
For more detail, see \cite[Section-IV]{oltdr}.

Consider now a truncated RSD in such a way that $\mu(x)=\sum_{d\le d_0}\mu_d x^d$, where $\lim_{n\to\infty}\frac{d_0}{n}=0$. 
Let $H$ be now an $m\times n$ matrix generated by a truncated $\mu(x)$.
The reason behind this truncation is that in practice,  most of the fractions  of the $\mu(x)$ in (\ref{eq:grsd}) are too small to get $ m\mu_d \ge 1$. Fractions for higher degrees however  should be assigned appropriately to meet the constraint  on the density (in number of nonzero entries of $H$),
\begin{equation}
a_r=\sum_{d\le d_0} d\cdot \mu_d  \quad \ge \quad \frac{\ln(n/\epsilon )}{(1+\gamma)}, \label{eq:avg}
\end{equation}
where $a_r$ is the average row-degree of $H$. 
By doing so, all columns of $H$ of \sysref{sys:ini} are nonzero with probability at least $1-\epsilon$.
This constraint can be explained by looking at the column-degree distribution of $H$,
\begin{equation}
  \lambda(x)=\prod_{d=1}^{d_0} \left[\left(1-\frac{d}{n}\right)+\left(\frac{d}{n}\right)x\right]^{(1+\gamma)n \mu_d}, \label{eq:lambda}
\end{equation} 
as follows.
From $\lambda(x)$, we have  $n\lambda(0)\approx ne^{-(1+\gamma)a_r}$ as the expected number of null columns of $H$. 
With an appropriate $d_0$, therefore,  fractions of $\mu(x)$ should be assigned to meet the inequality $ne^{-(1+\gamma)a_r} \le \epsilon <1$, that is equivalent to (\ref{eq:avg}).

For short $n$, even if a truncated $\mu(x)$ meets the constraint (\ref{eq:avg}), experiments exhibited that the DEP of codes  by $\mu(x)$  alone exhibits error-floors over a large range of $\gamma$.
 A desirable feature of $\mu(x)$ is however that, for $\gamma \ge 0$, the DEP   is  mainly contributed by the rank-deficient probabilities  $\Pr(\text{Rank}(H)=n-\eta)$ of  small $\eta$.
In \figref{fig:a}, for example,  the  curve $\epr(1+\gamma,100,\mu)$, where  $\mu(x)$ is a truncated one in \tabref{tab:rho}, is the DEP approximated by the GE  in \cite{oltdr} called  the Separated MLDA (S-MLDA). As can be seen clearly, it never reaches the bound $\delta=10^{-3}$ for $1+\gamma \le 1.3$.  
As $\gamma$ increases, on the other hand, it is very close to the deficiency curve $\eta=1$ and  is almost identical to the sum of deficiency probabilities of $\eta=1,\dots,7$.  
These deficiency probabilities  can be lowered to  near zero dramatically when  enough number of dense rows are supplemented to $H$.
The portion of the density increased by  the dense rows alone however could be much larger than expected.  
Therefore, with a desired $\delta$, the fraction for $\rho_{n/2}$ of $\rho(x)$ in (\ref{eq:rho1}) should be as  small as possible, while maintaining the $\gamma_{\min}(\delta,n,\rho)$ near the $\gamma_K(\delta,n)$.

Let $\mcd_1=\{1,2,\dots, d_s \}$, and  let $\mcd_2$ be a set of few spike  degrees $d$ such that $d_s < d \le d_0$ for some $d_0$. 
A truncation of $\mu(x)$ is summarized as follows.  
\begin{enumerate}[\; R1)]
 \item Generate the  $\mu(x)$  in (\ref{eq:grsd}) with  desired $S$, $n$, and $\epsilon$. 
 \item  Take  a few spike terms for $\mcd_2$, if necessary, such that $\sum_{d\in\mcd_2}\mu_d = 1- \sum_{d\in \mcd_1}\mu_d$, and   at the same time to hold  $a_r \ge \frac{\ln(n/\epsilon)}{1+\gamma}$ as in (\ref{eq:avg}).
\end{enumerate}
 Thus, hopefully, columns of $H$ by a truncated $\mu(x)$ have a  one in some rows  of degree $d \in \mcd_1 \cup \mcd_2$ with probability at least $1-\epsilon$.
An exemplary $\mu(x)$  generated by  R1 and R2 is listed in \tabref{tab:rho}, and its supplementation $\rho(x)$ with  various fractions for $\rho_{n/2}$ was used for computer simulations presented in  \figref{fig:b} and \figref{fig:sperflt}.

\section{The Proposed Estimate for DEP of LT Codes \label{testimate}}\noindent%

We first introduce the optimization of $\rho(x)$ in \cite{oltdr} that uses estimated UBDEP's.
Let $H$ of \sysref{sys:ini} be  generated by a supplementation $\rho(x)$ in (\ref{eq:rho1}) with $m=(1+\gamma)n$. 
By rearranging rows,  we have $H=\left[\begin{smallmatrix} \tilde{H} \\ \hat{H} \end{smallmatrix} \right]$, where  the sparse part $\tilde{H}$ is generated by a truncated $\mu(x)$ and the dense part $\hat{H}$ is formed by random rows of degree $\frac{n}{2}$.
Let $B(m,k,p)=\binom{m}{k}p^k (1-p)^{m-k}$ be the Bernouli probability with $0\le p\le 1$. 
Assume that the dense part $\hat{H}$ attains $k$ rows with probability $B(m,k,\rho_{n/2})$. 
With the sparse part $\tilde{H}$, let $\varphi_{\mu}(\eta)=\Pr(\text{Rank}(\tilde{H})=n-\eta)$. 
Let $\vartheta_{\rho} (k,\eta)=\Pr(\text{Rank}(H)<n|k,\eta)$  the conditional probability that $\text{Rank}(H)<n$, given that $\hat{H}$ attained $k$ rows and $\text{Rank}(\tilde{H})=n-\eta$. 
Finally, let $\vartheta_{\rho} (k)=\Pr(\text{Rank}(H)<n|k)$ the conditional probability that $\text{Rank}(H)<n$ given that $\hat{H}$ attained $k$ rows. 
We have  $\vartheta_\rho(k)=\sum_{\eta=1}^{n} \vartheta_\rho(k,\eta)\cdot \varphi_{\mu}(\eta)$.

The UBDEP in \cite{oltdr} was formulated in two steps: first by expressing the DEP in (\ref{eq:dfr2}) as the sum 
  \begin{equation}
  \epr(1+\gamma,n,\rho) = \sum_{k=0}^{m}B(m,k,\rho_{n/2})\cdot \vartheta_{\rho} (k), \label{eq:dfr1}
 \end{equation} 
second, by finding an upper-bound for $\vartheta_{\rho} (k)$ as shown in the following theorem. (For the proof, see \cite[{\bf Theorem~IV}]{oltdr}).
\begin{thm1}[The UBDEP of LT codes  by $\rho(x)$]\label{thm:upmu} 
 Since $\vartheta_{\rho} (k,\eta)\le \frac{1}{2^{k-\eta}}$ for $1\le\eta\le k$ by the union bound and $\vartheta_{\rho} (k,\eta)=1$ for $\eta>k$, we have
\begin{equation}
\vartheta_{\rho} (k) \le \left( \overline{\vartheta_{\rho} (k)}=\sum_{\eta=1}^{k} \frac{\varphi_\mu(\eta)}{2^{k-\eta}} + \sum_{\eta >k}\varphi_\mu(\eta)\right). \label{eq:cdrdef}
\end{equation}
This yields the UBDEP as shown below
\begin{equation}
 \epr(1+\gamma,n,\rho) \le \sum_{k=0}^{m}B(m,k,\rho_{n/2})\cdot \overline{\vartheta_{\rho} (k)}. \label{eq:udep}
\end{equation}
\end{thm1}

\begin{figure}[ht]
\centering
\fbox{\epsfig{file=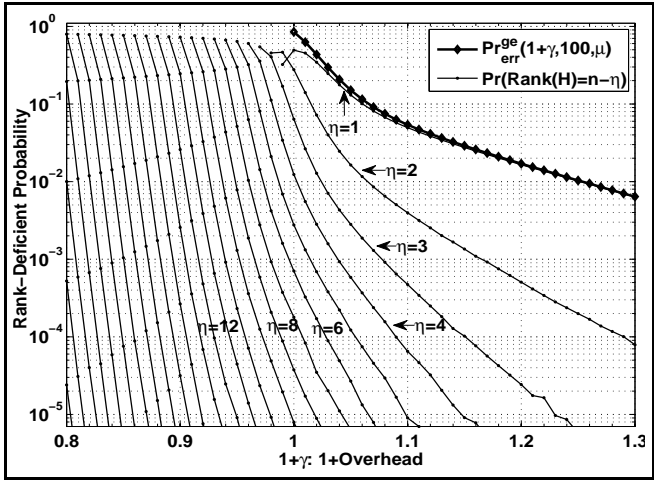,width=3.3in, height=2.4in}}
\caption{
The curves $\eta=1,2,\dots,$  represent $\phi_\mu(\eta)=\Pr(\text{Rank}(\tilde{H})=n-\eta)$  approximated by the S-MLDA, where $\tilde{H}$ is generated by the $\mu(x)$ in \tabref{tab:rho} with $n=100$. 
\label{fig:a}}
\end{figure}

Notice that, once the deficient probabilities $\phi_\mu(\eta)$, $\eta=1,2,\dots,\eta_0$, are estimated   for some $\eta_0$ (e.g., the deficiency curves $\eta=1,2,\dots,27$ in \figref{fig:a}), the UBDEP in (\ref{eq:udep})  can be estimated  very fast for any fraction for $\rho_{n/2}$.  Furthermore, experiments exhibited that the estimate  is also close to the DEP approximated by GE over \sysref{sys:ini}. 
Thus the overall shape of DEP including its error-floor region is predictable from the estimates right away.  
Exemplary optimizations using these estimates are presented in \cite{oltdr}.

We shall now decompose the $\Pr(\text{Rank}(H)=n)$  in (\ref{eq:dfr}) as a sum of conditional full-rank probabilities of $H$. 
Let us clarify some notations first. 
With $0\le\omega \le \min\{k,\eta\}$, let 
\begin{equation}
\zeta(\eta,\omega,k)=\Pr(\text{Rank}(H)=(n-\eta)+\omega|\eta,k),\label {eq:zetak1}
\end{equation}
 the conditional probability that $\text{Rank}(H)=(n-\eta)+\omega$ given that $\text{Rank}(\tilde{H})=n-\eta$ and $\hat{H}$ attained $k$ rows. 
Assume that the dense part $\hat{H}$ attains $k$ rows with probability $B(m,k,\rho_{n/2})$.
Let $\Pr(\text{Rank}(H)=n|k)$ denote the conditional full-rank probability given that $\hat{H}$ attained $k$ rows.
We  have, first, 
\begin{equation}
  \Pr(\text{Rank}(H)=n|k)= \sum_{\eta=0}^{k} \zeta(\eta,\eta,k)\phi_\mu(\eta), \label{eq:cfrankk}
\end{equation}
where $\phi_\mu(\eta)=\Pr(\text{Rank}(\tilde{H})=n-\eta)$. Second,
\begin{eqnarray}
\Pr(\text{Rank}(H)=n) &=& \sum_{k=0}^{m} B(m,k,\rho_{n/2})\nonumber \\
                      & & \; \cdot \Pr(\text{Rank}(H)=n|k). \label{eq:frp0}
 \end{eqnarray} 
Then finally, we have
\begin{equation}
  \epr(1+\gamma,n,\rho) = 1-  (\text{\ref{eq:frp0}}). \label{eq:frp}
\end{equation}

An explicit formulation of $\zeta(\eta,\omega,k)$ in (\ref{eq:cfrankk}) is possible by interpreting Kovalenko's rank-distribution \cite{kv1,kv2,cc,rg,allerton07} as the conditional one as shown in the following lemma.
\begin{lem}[The Conditional  Kovalenko's Rank-Distribution] \label{lm:gkrd} 
For any $(\eta,\omega,k)$ with $\omega\le \min\{\eta,k\}$, we have
 \begin{equation}
\zeta (\eta,\omega,k) = \frac{S(\omega,l)}{2^{l(\eta-\omega)}}\prod_{i=1}^{\omega}\left(1-\frac{1}{2^{\eta+1-i}} \right), \label{eq:zetak}
\end{equation}
where, with $l=k-\omega$, 
\begin{equation}
 S(\omega,l)=\sum_{i_1=0}^{\omega} \sum_{i_2=i_1}^{\omega}\cdots  \sum_{i_l=i_{l-1}}^{\omega}\frac{1}{2^{i_1+\cdots+i_l}}, \label{eq:S}
\end{equation}
holding the recursion 
\begin{equation}
 S(\omega,l)=\frac{1}{2^l}S(\omega-1,l) + S(\omega,l-1). \label{eq:SR}
\end{equation}
\end{lem}
\begin{IEEEproof}
The proof can be accomplished  inductively by using the recursions  (\ref{eq:SR}) and
\begin{eqnarray}
 \zeta(\eta, \omega,k+1)&=& \zeta(\eta,\omega-1,k)\left(1-\frac{1}{2^{\eta-\omega+1}} \right)\nonumber \\                                       & &  + \quad \zeta(\eta,\omega,k)\frac{1}{2^{\eta-\omega}}. \label{eq:zetarc}
\end{eqnarray}
For detailed proof, we refer readers to \cite[ {\bf Lemma~IV.3}]{allerton07}. 
\end{IEEEproof}

\begin{table}[t]
\begin{center}
\begin{tabular}{|c|l|}
 \hline
                 & $(\mu_d)_{d=1}^{5}=(0.014,0.481,0.152,0.082,0.048)$ \\   
$\mathcal{D}_1$  & $(\mu_d)_{d=6}^{10}=(0.034,0.024,0.024,0.012,0.012)$\\ 
 \hline
$\mathcal{D}_2$  & $\mu_{25}=0.059$, $\mu_{35}=0.058$ \\ 
 \hline
\end{tabular}
\end{center}
\caption{Fractions of $\mu(x)$  generated by the steps R1) and R2) with $S=10$ and $n=10^3$. \label{tab:rho}}
\end{table}

\begin{figure}[ht]
\centering
\fbox{\epsfig{file=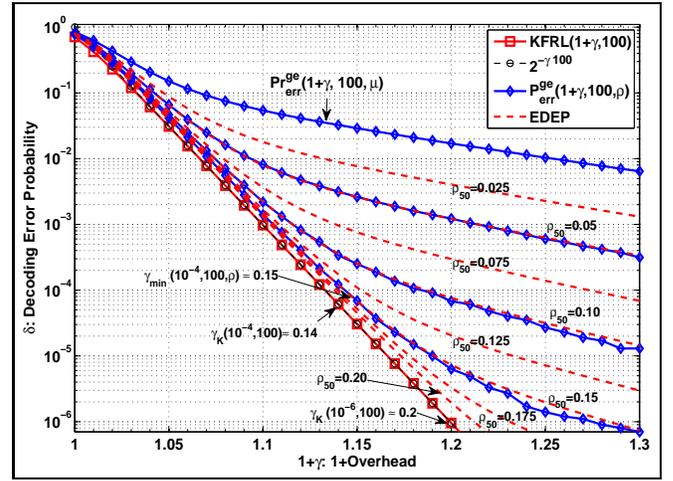,width=3.3in, height=2.4in}}
\caption{ \label{fig:b} In the figure, $\rho(x)$ is a supplementation of the $\mu(x)$ in \tabref{tab:rho} with an assigned dense  fraction for $\rho_{n/2}$. 
For each fraction for $\rho_{n/2}$, the DEP curve  is approximated by the S-MLDA and  the EDEP  is computed by using the finite version in (\ref{eq:efrp}).}
\end{figure}

\begin{thm1}\label{thm:dfr}
Assume that $\phi_\mu (\eta)$ in (\ref{eq:cfrankk}) are explicitly known over a deficiency range $1\le\eta\le \eta_0 $ for some $\eta_0$. Then the DEP  in (\ref{eq:frp}) is  explicitly computable.
\end{thm1}
\begin{IEEEproof}
Since  $\zeta(\eta,\omega,k)$ in (\ref{eq:zetak1}) is  explicitly computable  by  the recursions (\ref{eq:SR}) and (\ref{eq:zetarc}),   $\zeta(\eta,\eta,k)$ in (\ref{eq:cfrankk}) is computable by $S(\eta, k-\eta)\prod_{i=1}^{\eta}(1-2^{-i})$. Therefore, the DEP in (\ref{eq:frp}) is  explicitly computable.
\end{IEEEproof}

  The  DEP in (\ref{eq:frp}) is very practical  in two respects. 
First, for any fraction for $\rho_{n/2}$,  experiments exhibited that the EDEP   is almost identical to the  DEP approximated by GE.  
Second,  the full-rank probability in (\ref{eq:frp0}) can be estimated  very rapidly, and  therefore, a fine optimization of $\rho(x)$ is obtainable very fast by checking EDEP's with  various fractions for $\rho_{n/2}$.
The following example shows the viability that the EDEP is very close to the DEP approximated by the S-MLDA over \sysref{sys:ini}.  An exemplary optimization of $\rho(x)$ using EDEP's is also presented in the example.

\begin{eg1}
In \figref{fig:a},  the deficiency curves  $\eta=1,2,3,\dots,27$,  represent  $\phi_\mu(\eta)=\Pr(\text{Rank}(\tilde{H})=n-\eta)$ approximated by  the S-MLDA over \sysref{sys:ini}.  
In \figref{fig:b}, blue curves  are DEP's approximated  by the S-MLDA, where $\rho(x)$ is a supplementation of the $\mu(x)$ in \tabref{tab:rho} with a dense fraction in $\rho_{n/2}=\{0.05,0.10, 0.15\}$, and
dashed curves in red  are EDEP's of codes by $\rho(x)$ of having a dense fraction in $\rho_{n/2}=\{0.025,0.05,\dots,0.175,0.20 \}$ computed by using  a truncated version of (\ref{eq:frp0}),
\begin{equation}
 \text{(\ref{eq:frp0})} :=\sum_{k=l-5}^{l+5} B(m,k,\rho_{n/2}) \Pr(\text{Rank}(H)=n|k), \label{eq:efrp}
\end{equation}
where $l=\left\lfloor m\rho_{\frac{n}{2}}\right\rfloor$.
Notice in \figref{fig:b} that, for each assigned fraction for $\rho_{n/2}$, the EDEP by (\ref{eq:efrp}) above is almost identical to the blue one approximated by the S-MLDA.

Let  $\delta=10^{-4}$ be a given error-bound. Notice from the graph of $K(1+\gamma,100)$ that $\gamma_K(10^{-4},100):=0.14$.
Assume that we want a  $\rho(x)$ such that $0.14 \le \gamma_{\min}(10^{-4},\rho,100) \le 0.15$.
By checking EDEP's with various fractions for $\rho_{n/2}$, we see that the dense fraction should be larger than $0.125$, but the fraction $\rho_{n/2}= 0.15$  is large  enough for the optimal $\rho(x)$. 
With $\delta=10^{-6}$ and $\gamma_K(10^{-6},100)=0.2$, similarly, we get $\rho_{n/2}\approx 0.20$ for the constraint $0.2 \le \gamma_{\min}(10^{-6},\rho,100) \le 0.21$.
\hfill \IEEEQEDclosed
\end{eg1}

The KFRL in \cite{kfrl} can be explained by a particular case of $\zeta(\eta,\omega,k)$ in (\ref{eq:zetak}). 
To see this, let $H=\hat{H}$ so that  $\tilde{H}=\emptyset$.  
By replacing the $k$  with $n+k$, $\eta$ with $n$, and $\omega$ with  $n-s$ (hence $l=k+s$) in (\ref{eq:zetak}), we have a finite version of Kovalenko's rank distribution with $q=2$ as shown below
\begin{equation}
\Pr(\text{Rank}(\hat{H})=n-s) =\frac{S(n-s,k+s)}{2^{(k+s)s}} \prod_{i=s+1}^{n}\left(1-\frac{1}{2^i} \right).
\end{equation} 
Since $\lim_{n\to\infty}S(n-s,k+s)=\prod_{i=1}^{k+s}\left(1-\frac{1}{2^i} \right)^{-1}$ and the sequence $\{ S(n-s,k+s)\}_{n=s}^{\infty}$ is increasing, we have
\begin{equation}
\Pr(\text{Rank}(\hat{H})=n-s) \le \frac{1}{2^{s(k+s)}} \frac{\prod_{i=s+1}^{n}\left(1-\frac{1}{2^i} \right)}{\prod_{i=1}^{k+s}\left(1-\frac{1}{2^i} \right)}. \label{eq:n-s}
\end{equation}
The KFRL in (\ref{eq:kfrl0}) is then a particular case of the upper-bound above with $s=0$.
Observe in \figref{fig:b} that  as $\rho_{n/2}$ increases the DEP approaches closer to the  limit $K(1+\gamma,100)$.
Notice that the KFRL  is almost identical to $2^{-\gamma\cdot 100}$ as $\gamma$ increases.

%
%
\section{Further Experimental Results} \label{simulation}


\begin{figure}[t]
 \centerline{\fbox{\epsfig{file=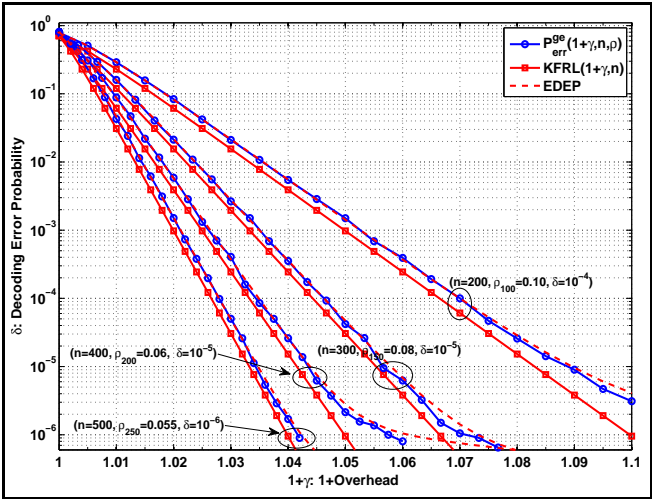,  width=3.3in, height=2.5in}}}
 \caption{Curves in the figure represent EDEP (dashed in red) by (\ref{eq:frp}), DEP (blue) by the S-MLDA,  and KFRL (red)  with destined error-bounds $\delta=10^{-4}$ for $n=200$, $10^{-5}$ for $n=300,400$, and $10^{-6}$ for $n=500$. }  \label{fig:sperflt}
\end{figure}

In this section, experimental results which show the viability of the EDEP for other block-lengths are presented.  
The  truncated RSD in \tabref{tab:rho} was used for the experiments with the block lengths $n=200,300,400,500$.

By using EDEP's with various fractions for $\rho_{n/2}$, we first investigated triple pairs of $(n,\rho_{n/2},\delta)$ for an optimal $\rho(x)$ in advance  as shown in \figref{fig:sperflt}, where $\delta$ is a destined error-bound.
With $(500,0.055,10^{-6})$, for example, the fraction $\rho_{250}=0.055$ is large enough for the supplementation $\rho(x)$, achieving the MSO $\gamma_{\min}(10^{-6},500,\rho)$ that is near the KFRO $\gamma_K(10^{-6},500)\approx 0.04$.
Then with each optimized $\rho(x)$, we approximated the DEP by applying the S-MLDA over \sysref{sys:ini}.
As can be seen clearly, each EDEP by $\rho(x)$ is almost identical to the DEP  approximated by the S-MLDA, and also, it is very close to the limit $K(1+\gamma, n)$ up to a destined error-bound $\delta$.

\begin{figure}[t] 
 \centerline{\fbox{\epsfig{file=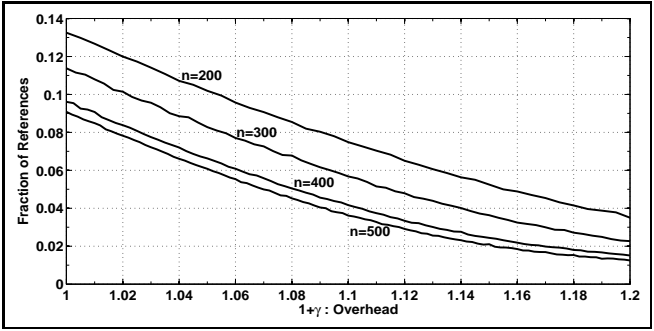,  width=3.3in,height=4.1cm}}}
 \caption{Curves represent the fractions $\epsilon=\frac{r}{n}$ with $n=200,300,400,500$ by the diagonal extension process over an $H$ by $\rho(x)$ used in \figref{fig:sreflt}}\label{fig:sreflt} 
\end{figure}

Let us now discuss the efficiency of LT decoding under the S-MLDA. 
To solve \sysref{sys:ini}, like other  MLDA's in \cite{eeldpc,bnm1,usp1}, the S-MLDA uses an approximate lower-triangulation of $H$ in such a way that $\bar{H}=PHQ^T=\left[ \begin{smallmatrix}A & B \\ C & D \end{smallmatrix}\right]$, where  $(P,Q)$ is a pair of row and column permutations obtainable by  the diagonal extension process in \cite{eeldpc,bnm1,oltdr}, and the right-top block $B$ is an $l\times l$ lower triangular matrix with $l=n-r$ close to $n$. 
The successful decoding by MPA in \cite{tornado1} is a particular case when  $\left[ \begin{smallmatrix}A  \\ C \end{smallmatrix}\right]=\emptyset$. 
If $\left[ \begin{smallmatrix}A  \\ C \end{smallmatrix}\right]\neq \emptyset$ then the S-MLDA transforms $\bar{H}$ to $\left[\begin{smallmatrix}\bar{A} & I_{l\times l} \\ \bar{C} & 0 \end{smallmatrix}\right]$ by pivoting columns of the  right-block $\left[\begin{smallmatrix} B \\ D \end{smallmatrix}\right]$ from the first to the last column of it, and then transforms the $r\times (r+\gamma n)$ block $\bar{C}$ to $\left[\begin{smallmatrix} I_{r\times r} \\ 0 \end{smallmatrix}\right]$ by a conventional GE.

Let $r=\epsilon n$. Since $\bar{C}$ is not sparse in general, the decoding complexity of the transformation by the GE  is $O((\gamma\epsilon^2+\epsilon^3) n^3)$. 
Hence the overall complexity of  decoding by the MLDA's is dominated by either $O((\gamma\epsilon^2+\epsilon^3) n^3)$ or  the density $|H|$. 
Thus, although its overall complexity is $O(n^3)$, the efficiency of the LT decoding under the S-MLDA can be measured in terms of the fraction $\epsilon=\frac{r}{n}$, and this is particularly true for short $n$.

In \figref{fig:sreflt}, curves represent  the fraction $\epsilon=\frac{r}{n}$ obtained by the diagonal extension process on $H$ generated by the $\rho(x)$ used in \figref{fig:sperflt}.
When  $n=500$ and $1+\gamma=1.1$, for instance, the point $(1.1, 0.038)$ indicates  that, with a $550\times 500$ random $H$ by the $\rho(x)$ with $\rho_{n/2}=0.055$, the column-dimension of $\bar{C}$ is about $r=20$ that is much smaller than  $n=500$ the column-dimension of $H$.
This substantiates that decoding of the codes under the S-MLDA becomes very efficient as $\gamma$ increases.
 


%

%
%
%
%
\section{Summary} \label{conclusion}
    
In \secref{trsd}, a simple approach of generating a truncated RSD is presented.
In \secref{testimate}, explicit formulations of DEP and UBDEP are derived  and utilized for obtaining an optimal $\rho(x)$, and KFRL is explained as a particular case of the conditional Kovalenko's rank-distribution.
In  \secref{simulation}, experimental results  which show the viability of the EDEP  are presented.

\end{document}